\pdfoutput=1

\documentclass[journal]{IEEEtran}

\usepackage[utf8]{inputenc}
\usepackage{csquotes}
\usepackage[english]{babel}

\usepackage{xcolor}
\usepackage{graphicx}
\usepackage{epstopdf} 
\usepackage{ifpdf}
\ifpdf
  \DeclareGraphicsExtensions{.pdf,.png,.jpg,.eps}
\else
  \DeclareGraphicsExtensions{.eps,.pdf}
\fi

\usepackage{tikz}
\usetikzlibrary{shapes,arrows}
\usetikzlibrary{calc}

\usepackage{adjustbox} 

\usepackage{amsfonts,amssymb,amsmath,amsthm} 
\usepackage{fixmath} 				  
\usepackage{mathtools}
\usepackage{mathrsfs}
\usepackage{dsfont}
\usepackage{centernot}
\DeclareMathAlphabet{\mathpzc}{OT1}{pzc}{m}{it}
\usepackage{mathabx} 
\usepackage{textcomp} 

\usepackage{todonotes}

\usepackage{enumerate}
\usepackage{paralist}
\usepackage{longtable}
\usepackage{multirow}
\usepackage{comment}

\usepackage[natbib=true,backend=bibtex,style=ieee,
firstinits=true,eprint=false,doi=false,isbn=false,url=false,backref=false,texencoding=utf8,bibencoding=utf8]{biblatex}
\IfFileExists{./Library.bib}{

\addbibresource{./Library.bib}

}{

  \addbibresource[location=remote]{https://www.dropbox.com/s/r00o9lw76cksxvj/Library.bib?dl=1}

}

\usepackage{etoolbox}

\newtoggle{euler}
\togglefalse{euler}

\newcommand{\R}{\mathbb{R}}

\newcommand{\Ras}{\mathbb{R}_{\geq0}}

\iftoggle{euler}{%

}{

}

\newcommand{\cK}{\mathcal{K}}
\newcommand{\cL}{\mathcal{L}}

\theoremstyle{definition}
\newtheorem{defn}{Definition}

\newtheorem{assum}{Assumption}

\theoremstyle{remark}

\theoremstyle{plain}
\newtheorem{thm}[defn]{Theorem}

\newcommand{\sign}{\mathtt{sign}}
\newcommand{\cl}{\mathtt{cl}}
\newcommand{\Liploc}{\cL\mathpzc{ip}_\mathrm{loc}}
\DeclareMathOperator{\interior}{\mathtt{int}}
\newcommand{\id}{\mathtt{id}}
\renewcommand{\div}{\mathtt{div}}

\begin{document}

\title{Example Demonstrating the Application of Small-gain and Density Propagation Conditions for Interconnections}

\author{\IEEEauthorblockN{Humberto~Stein~Shiromoto\IEEEauthorrefmark{1}, Petro~Feketa\IEEEauthorrefmark{2}, Sergey~Dashkovskiy\IEEEauthorrefmark{3}}\\ \vspace{1em}%
	\small\IEEEauthorblockA{\IEEEauthorrefmark{1} The Australian Centre for Field Robotics.  The
Rose Street Building J04, The University of Sydney, NSW 2006, Australia}\\

 \IEEEauthorblockA{\IEEEauthorrefmark{2} Department of Civil Engineering, University of Applied Sciences Erfurt, Altonaer Str. 25, 99085 Erfurt, Germany}\\

\IEEEauthorblockA{\IEEEauthorrefmark{3} Institute of Mathematics, University of W\"{u}rzburg, Emil-Fischer-Str. 40, 97074 W\"{u}rzburg, Germany}

\thanks{This work was partially supported by the German Federal
Ministry of Education and Research (BMBF) as a part of the
research project ``LadeRamProdukt''. Email addresses: humberto.shiromoto@ieee.org
(Humberto Stein Shiromoto), petro.feketa@fh-erfurt.de
(Petro Feketa), sergey.dashkovskiy@uni-wuerzburg.de
(Sergey Dashkovskiy).}
}

 \maketitle

\begin{abstract}                          
This work provides an example that motivates and illustrates theoretical results
related to a combination of small-gain and density propagation conditions.
Namely, in case the small-gain fails to hold at certain points or intervals
the density propagation condition can be applied to assure global stability properties.
We repeat the theoretical results here and demonstrate how they can be applied in the proposed example.
\end{abstract}


\section{Introduction}\label{sec:introduction}

Here we will consider two nonlinear interconnected systems, each of them is ISS, and ask whether
the interconnection is ISS as well. Typically one uses the so-called small-gain condition
to assure this property. However in case this condition fails to hold at several points or intervals further conditions to assure global stability properties are necessary.
This condition can be written in terms of a density propagation inequality \cite{Angeli:2004}.
Moreover if one tries to verify the small-gain condition numerically it can easily happen
that this condition fails due to the unprecise computer calculations. Also in this case
the density propagation condition can help to fill such gaps.

The example considered below demonstrates this kind of situations and illustrates how a combination of small-gain and density propagation conditions can be applied.
First we recall the related theoretical results, that will be published elsewhere, then we
will consider the example in detail.
\section{Preliminaries and notation}\label{sec:notation}
The notation $\overline{\mathbb{N}}$ (resp. $\overline{\mathbb{R}}$) stands for the set $\mathbb{N}\cup\{\infty\}$ (resp. $\mathbb{R}\cup\{\infty\}$). For a given $a,b\in \overline{\mathbb{R}}$ let $\mathbb N_{[a,b]}=\{s\in\overline{\mathbb{N}}:a\leq s \leq b\}$. Let $\mathbf{S}\subset\R^n$, its closure (resp. interior) is denoted as $\cl\{\mathbf{S}\}$ (resp. $\interior\{\mathbf{S}\}$). We recall the following standard definitions: a function $\alpha:[0,\infty)\to[0,\infty)$ is of class $\mathcal K$ when $\alpha$ is continuous, strictly increasing, and $\alpha(0)=0$. If $\alpha$ is also unbounded, then we say it is of class $\mathcal K_\infty$. A continuous function $\beta:[0,\infty)\times[0,\infty)\to[0,\infty)$ is of class $\mathcal{KL}$, when $\beta(\cdot,t)$ is of class $\mathcal K$ for each fixed $t\geq 0$, and $\beta(r,t)$ decreases to $0$ as $t\to\infty$ for each fixed $r\geq 0$.

Consider the interconnection of two systems $\Sigma_1$ and $\Sigma_2$
\begin{equation}\label{eq:subsystem}
	\Sigma_i:\quad \dot{x}_i=f_i(x_1(t),x_{2}(t),u_i(t)),\quad i=1,2
\end{equation}
$x_i(t)\in\mathbb{R}^{n_i}$ is the state of $\Sigma_i$ and $u_i(t)\in\mathbb{R}^{m_i}$ is its external input, $f_i$ is assumed to be of class $\mathcal{C}^1$ and satisfy $f_i(0,0,0)=0$. This interconnection can be written as
\begin{equation}\label{newthree}
\dot x=f(x(t),u(t))
\end{equation}
with the state $x=(x_1,x_2)\in\mathbb{R}^n,\; n=n_1+n_2$, dynamics $f=(f_1,f_2)$ and input $u=(u_1,u_2)\in\mathbb{R}^m,\; m=m_1+m_2$.
\begin{defn}[{\cite{Sontag1989}}]
	The system \eqref{newthree} is called \emph{input-to-state stable} (ISS) if there exist functions $\beta\in\mathcal{K}\mathcal{L}$ and $\tilde{\gamma}_{u}\in\mathcal{K}_\infty$ such that, for each initial condition $x(0)$ and each measurable essentially bounded input $u(\cdot)$, the solution $x(\cdot)$ of \eqref{newthree} satisfies
	\begin{align*}
		|x(t)|\leq& \beta(|x(0)|,t)+\tilde{\gamma}_{u}(|u|_{\infty}) \quad \forall t\geq 0.
	\end{align*}
\end{defn}
ISS is equivalent to the existence of an ISS-Lyapunov function, which we define for each subsystem in \eqref{eq:subsystem}:
\begin{defn}
	A function $V_i\in\Liploc(\R^{n_i},\Ras)$ is called \emph{storage function} if for some $\underline{\alpha}_i,\overline{\alpha}_i\in\cK_\infty$ it holds that  $\underline{\alpha}_i(|x_i|)\leq V_i(x_i)\leq \overline{\alpha}_i(|x_i|)\quad\forall x_i\in\R^{n_i}$.
\end{defn}
\begin{defn}
	A storage function $V_i$ is called \emph{ISS-Lyapunov function for \eqref{eq:subsystem}} if for some $\gamma_{i,j},\gamma_{i},\alpha_i\in\cK_\infty$ the implication
	\begin{subequations}\label{eq:ISS Lyapunov}
	\begin{equation}\label{eq:ISS Lyapunov:if}
		V_i(x_i)\geq\max\bigg\{\gamma_{i,j}(V_{j}(x_{j})),\gamma_{i}(|u_i|)\bigg\}
	\end{equation}	
	\begin{equation}\label{eq:ISS Lyapunov:then}
	\Rightarrow\;\nabla V_i(x_i)\cdot f_i(x_i,x_{j},u_i)\leq-\alpha_i(|x_i|).
	\end{equation}
	\end{subequations}
  holds for each $x_i\in\mathbb R^{n_i}$, $x_j\in\mathbb R^{n_j}$, and $u_i\in\mathbb R^{m_i}$.
$\gamma_{i,j}$ (resp. $\gamma_{i}$) is called \emph{interconnecting} (resp. \emph{external}) \emph{ISS-Lyapunov gain}.
\end{defn}

Stability of the resulting interconnected system \eqref{newthree} can be deduced from the small-gain theorem \cite{Jiangetal:1996}: if the  interconnecting ISS-Lyapunov gains satisfy
\begin{equation}\label{eq:SGC}
\gamma_{12}\circ\gamma_{21}(s)<s\quad \forall s>0,
\end{equation}
then system \eqref{newthree} is input-to-state stable.

In this paper we assume that the graphs of $\gamma^{-1}_{12}$ and $\gamma_{21}$ have several points of intersection. It means that small-gain condition does not hold globally and the previously known results cannot be utilized to verify global stability properties of the interconnection. To guarantee the desired stability properties of the interconnection, the dual to Lyapunov's techniques \cite{Angeli:2004,Rantzer:2001} is employed in specific domains of the state space.
Our approach extends the results of \cite{SteinShiromoto2015} to the case of arbitrary number of intersection points of $\gamma^{-1}_{12}$ and $\gamma_{21}$ and allows for external inputs. Moreover our stability conditions are less restrictive then in \cite{SteinShiromoto2015}.

\begin{assum}\label{hyp:ISS}
Let for each $i=1,2$ there exist an ISS-Lyapunov function $V_i$ for $\Sigma_i$ from \eqref{eq:subsystem} with the corresponding gain functions $\gamma_{i,j},\gamma_{i}\in\cK_\infty$, and $\alpha_i\in\cK_\infty$.
\end{assum}
\begin{assum}\label{hyp:localSGC}
Let the intersection points of the graphs of $\gamma^{-1}_{12}$ and $\gamma_{21}$ be given. These points define a sequence of intervals $\mathbf{M}_k=(\underline{M}_k, \overline{M}_{k})$ where the small-gain condition \eqref{eq:SGC} holds for any $s\in\mathbf{M}_k$, $k\in\mathbb N_{[1,\ell]}$, $\ell\in{\mathbb{N}}$.
\end{assum}
For a given $\delta\in\mathbb R_{\geq 0}$ let $\mathbf{L}_i(\delta)=\{x_i\in \mathbb{R}^{n_i}:V_i(x_i)\leq \delta\}$.
\begin{thm}\label{prop:regional ISS}
Let Assumptions \ref{hyp:ISS} and \ref{hyp:localSGC} hold.
Then there exists $\gamma\in\mathcal K_{\infty}$ such that almost all solutions to system \eqref{newthree} starting in the set $\mathbf{B}_k$ converge to a neighborhood of the set $\mathbf{A}_k$ with radius $\gamma(|u|_{\infty})$, where
\begin{equation}\label{eq:Ak}
\begin{split}
\mathbf{A}_k=\{x\in \mathbb{R}^n:
x_1\in\mathbf{L}_1(\max\{\underline{M}_k, \gamma_{12}(\underline{M}_k)\}),\\ x_2\in\mathbf{L}_2(\max\{\gamma_{21}(\underline{M}_k),\gamma_{21}\circ\gamma_{21}(\underline{M}_k)\})\},
\end{split}
\end{equation}
\begin{equation}\label{eq:Bk}
\mathbf{B}_k=\{x\in \mathbb{R}^n:x_1\in\mathbf{L}_1(\overline{M}_k), x_2\in\mathbf{L}_2(\gamma_{21}(\overline{M}_k))\}.
\end{equation}
If $\overline M_\ell\neq \infty$ then the above mentioned convergence holds only for some bounded inputs $u$.
\end{thm}
The input-to-state stability of system \eqref{newthree} follows trivially when $\ell=1$, $\underline M_1=0$, $\overline M_1=\infty$; then $\mathbf{A}_1=\{0\}$ and $\mathbf{B}_1= \mathbb{R}^2$. However, when $\ell>1$, solutions to \eqref{newthree} starting in the set $\mathbf{A}_k$ may converge to an $\omega$-limit set  \cite[Birkhoff's Theorem]{Isidori:1995} that lies inside the set $\mathbf{A}_k$ and do not converge to a ball centred at the origin whose radius is proportional to the norm of the input. Due to this fact, the next assumption is needed to check the  asymptotic behaviour of solutions inside the sets $\mathbf{A}_k$. Let $\mathbf{B}_0=\emptyset$ and $\mathbf{A}_{\ell+1}=\mathbb R^n$.

\begin{assum}\label{hyp:a.e. dissipation inequality}
	Let for each $k\in \mathbb{N}_{[1,\ell+1]}$, there exist an open set
	$\mathbf{D}_k\subset\mathbb{R}^n$ satisfying $\{\mathbf{A}_k\setminus \mathbf{B}_{k-1}\}\subsetneq\cl\{\mathbf{D}_k\}$ and 
	\begin{itemize}
		\item A differentiable function $\rho_k:\mathbf{D}_k\to\mathbb{R}_{>0}$;
		
		\item A continuous function $q_k:\mathbf{D}_k\to\mathbb{R}_{\geq0}$ such that, for almost every $x\in\mathbf{D}_k$, $q_k(x)>0$;
		
		\item A function $\gamma_k\in\mathcal{K}$ such that, for every $x\in\mathbf{D}_k$ and for every $u\in\mathbb R^m$, the following implication holds
		\begin{subequations}
			\begin{equation}\label{eq:}
				\max_{i=1,2}V_i(x_i)\geq\gamma_k(|u|)\quad \Rightarrow
			\end{equation}
			\begin{equation}\label{dpi}
				\div(\rho_k f)(x,u):=\sum_{j=1}^n\tfrac{\partial (\rho_k f_j)}{\partial x_j}(x,u)\geq q_k(x)
			\end{equation}
		\end{subequations}
	\end{itemize}
\end{assum}
\begin{defn}\cite{Angeli:2004}
	The {origin} is called \emph{almost ISS for \eqref{newthree}} if it is locally asymptotically stable and for some $\gamma\in\mathcal{K}_\infty$
	\begin{equation*}
		\limsup_{t\to\infty}|x(t,x(0),u)|\leq\gamma(|u|_\infty)
	\end{equation*}
	holds for every input $u\in\mathcal{L}_{\mathrm{loc}}^\infty(\mathbb{R}_{\geq0},\R^m)$ and for almost every initial condition $x(0)\in \mathbb{R}^n$.
\end{defn}
\begin{thm}\label{thm:main result}
	Under Assumptions \ref{hyp:ISS}, \ref{hyp:localSGC}, and \ref{hyp:a.e. dissipation inequality} system \eqref{newthree} is almost input-to-state stable.
\end{thm}
\section{Illustrative example}
\begin{figure}[htbp!]
	\centering
	\includegraphics[width=\linewidth]{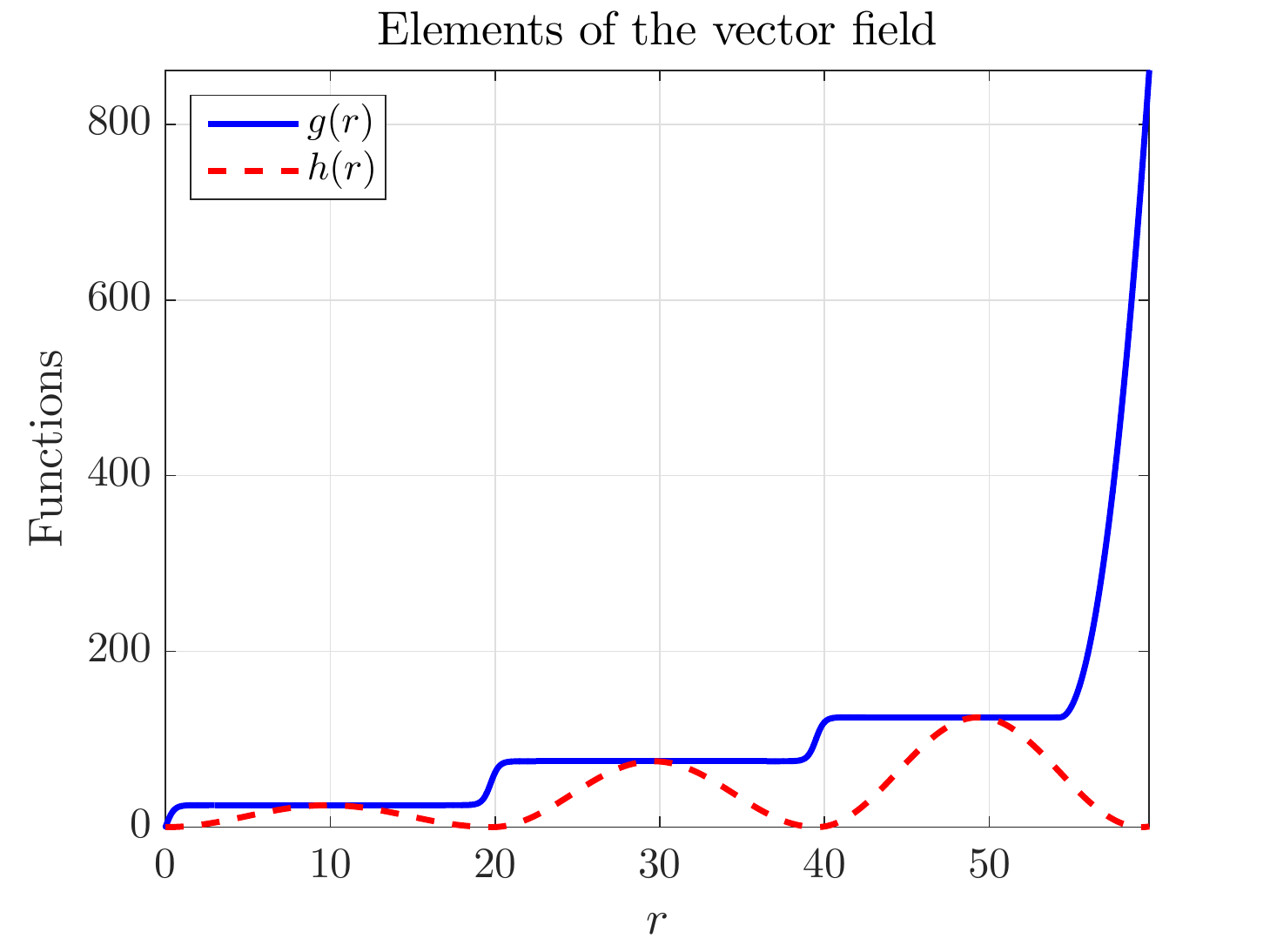}
	\caption{Plot of the functions $g$ and $h$, for $n=2$, and on the interval $[0,1.1\times11\pi^2/2]$.}
	\label{fig:components}
\end{figure}

Let $n\in\mathbb{N}\cup\{\infty\}$ be a constant value. Define the value $a=(4\pi^2n+3\pi^2)/2$ and the functions $g:\mathbb{R}\to\mathbb{R}$ as
\begin{subequations}
	\begin{align*}
		g(r)=&\ \tanh(2r)+\displaystyle\sum_{i=1}^n \sign(r)\\
		&\cdot(1 + \sign(r)\tanh(2(r - \sign(r)2\pi^2i))),
	\end{align*}
	for every $|r|\leq a$ and
	\begin{equation*}
	g(r)=\sign(r)((2n + 1) + (r - \sign(r)a)^2),
\end{equation*}
for every $|r|> a$, 
\end{subequations}
and $h:\mathbb{R}\to\mathbb{R}$ as
\begin{equation*}
	h(r)=\sin^2\left(\frac{r}{2\pi}\right)+\displaystyle\sum_{i=1}^n (\tanh(r - 2\pi^2i) + 1)\sin^2\left(\frac{r}{2\pi}\right).
\end{equation*}
Functions $g$ and $h$ are plotted in Figure \ref{fig:components}.

Although the equation $\tanh(r)-1=0$ has no (finite) solution. For computer programs running with a minimum precision $p$ for number representation there exist values $r>0$ such that, whenever the inequality $|\tanh(r)-1|<p$ is true, this value is rounded to zero. This is denoted as $\mathbin{\mathtt{round}}(\tanh(r)-1)=0$. The values where $r$ satisfies the equation $\mathbin{\mathtt{round}}(\tanh(r)-1)=0$ are said to be regions where $g$ is \emph{numerically constant}.

For each index $i=1,2$, consider the system defined by
\begin{equation}\label{eq:subsystem:example}
	\dot{x}_i=-\left(25+\frac{|u_i|_\infty}{a+1}\right)g(x_i)+25h(x_{3-i})+\left( \frac{|u_i|_\infty}{a+1}\right)^2.
\end{equation}
Denote the vector field describing the differential equation \eqref{eq:subsystem:example} by $f_i$ and let $f=(f_1,f_2)^\top$. This system is ISS, as shown below.

Letting either $|u_i|_\infty=0$ or $n=\infty$, Equation \eqref{eq:subsystem:example} reduces to
\begin{equation}\label{eq:subsystem:example:u=0}
	\dot{x}_i=-25g(x_i)+25h(x_{3-i}).
\end{equation}

For values of $a$ that are big enough (even infinite), due to numerical imprecision, the system resulting from the composition of \eqref{eq:subsystem:example} shows the presence of (possibliy infinitely) $(n+1)$ equilibrium points. These points are the maxima of the function $h$, namely, the points $r_k=2\pi^2k+\pi^2$, where $k\in\mathbb{Z}$.

\begin{figure}[htbp!]
	\centering
	\includegraphics[width=\linewidth]{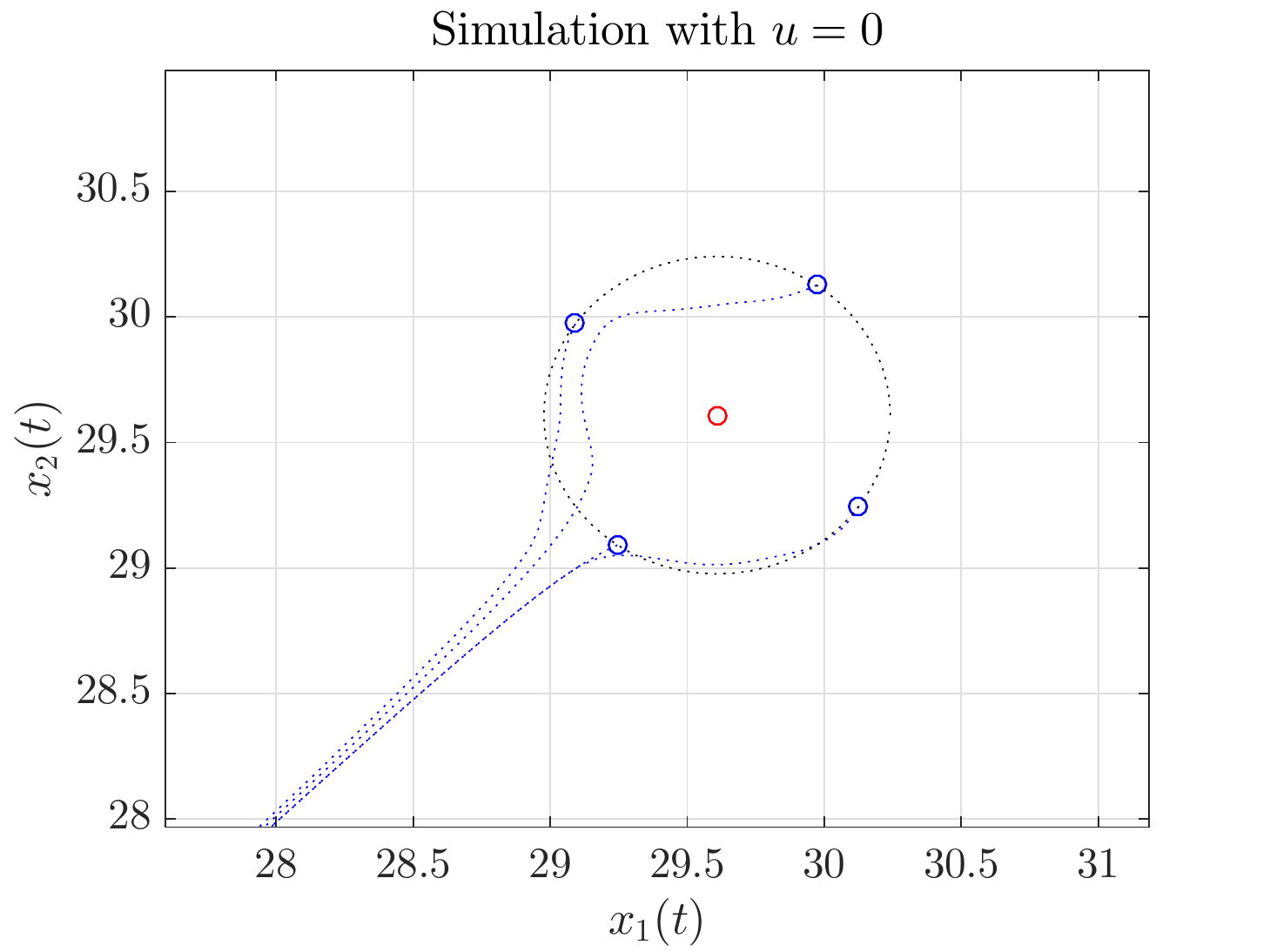}
	\caption{Simulation of the resulting interconnected system composed of  \eqref{eq:subsystem:example:u=0}. The circles are initial conditions. The red one is the equilibrium, i.e., the point $(x_1,x_2)=(r_1,r_1)$.}
	\label{fig:sim u=0}
\end{figure}

\begin{figure}[htbp!]
	\centering
	\includegraphics[width=\linewidth]{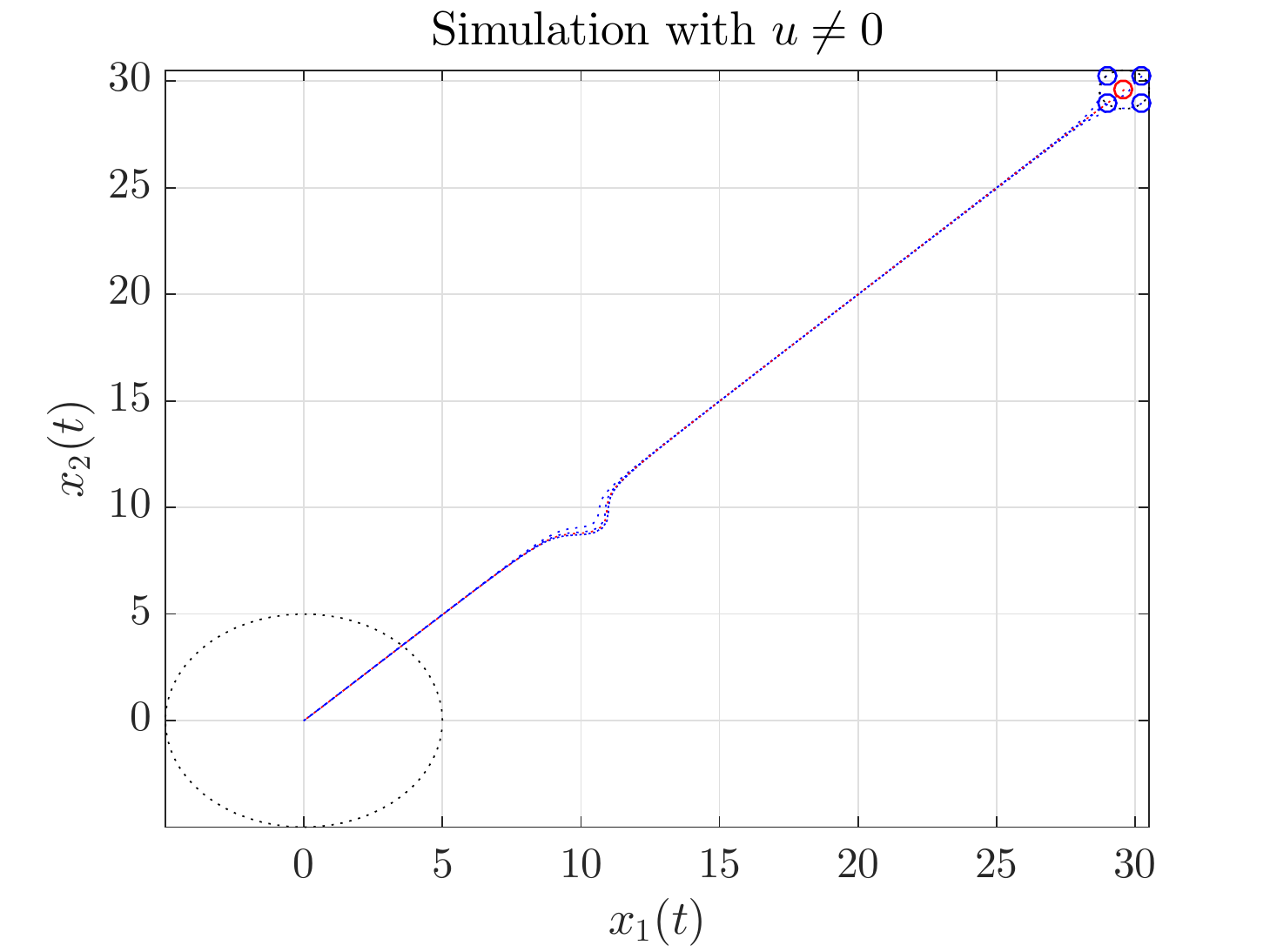}
	\caption{Simulation of the resulting interconnected system composed of  \eqref{eq:subsystem:example} with $|u_1|_\infty=3$ and $|u_2|_\infty=4$. The small circles are initial conditions. The large black circumference is the set $|(x_1,x_2)|=5$.}
	\label{fig:sim u!=0}
\end{figure}

Figures \ref{fig:sim u=0} and \ref{fig:sim u!=0} show a simulation of the system resulting interconnected system composed of  \eqref{eq:subsystem:example} without inputs and with nonzero inputs, respectively.

Note that the method presented in \cite{SteinShiromoto2015} can be employed to deduce the stability of the interconnected system, when the inputs are identically zero and $n=0$. However, when this is not the case, the method presented in this paper is needed. 

In the next paragraphs, the analysis is divided into two regions, according to whether the function $g$ is numerically constant or not.

\subsection{Proof that system \eqref{eq:subsystem:example} is ISS}

To see that system \eqref{eq:subsystem:example} is ISS, for each index $i=1,2$, define for every $x_i\in\mathbb{R}$, the function $V_i(x_i)=|x_i|$. The time derivative of this function along solutions to \eqref{eq:subsystem:example}, for every $(x_{3-i},u_i)\in\mathbb{R}\times\mathbb{R}$, the following inequality
\begin{align}
	\dot{V}_i(x_i)\leq&\sign(x_i)\dot{x}_i\nonumber\\
	\leq &-25g(V_i(x_i))+25h(V_{3-i}(x_{3-i}))\nonumber\\
	&+\left(\tfrac{|u_i|_\infty}{a+1}\right)^2.\label{eq:Vdot:tanh=1}
\end{align}

From here, two cases are analyzed, according to $n$ be finite or not.

Let $n$ be finite. Since the function $|g|$ positive definite, there exist functions $\alpha_i,\alpha_{i,u}\in\mathcal{K}$ such that, for every $x_i\in\mathbb{R}$, the inequality
\begin{equation}\label{eq:local ISS}
	\dot{V}_i(x_i)\leq -\alpha_i(V_i(x_i))+25h(V_{3-i}(x_{3-i}))+\alpha_{i,u}(|u_i|),
\end{equation}
holds, for every $(x_i,x_{3-i},u_i)\in\mathbb{R}\times\mathbb{R}\times\mathbb{R}$.

Thus, system \eqref{eq:subsystem:example} is ISS. Hence, Assumption \ref{hyp:ISS} holds.

\subsection{Proof that it satisfies small-gain conditions in the region where $g$ is not numerically constant}

Assume that $n$ is finite. This implies that there exists $n+2$ regions where the function $\mathbb{R}_{\geq0}\ni r\mapsto g(r)$ is not numerically constant. These regions are intervals $(\underline{M}_k,\overline{M}_k)$ of $\mathbb{R}_{\geq0}$ where the function $g$ is strictly numerically increasing, i.e., given two values $r_1,r_2\in(\underline{M}_k,\overline{M}_k)$ with $r_1<r_2$, $\mathbin{\mathtt{round}}(|g(r_1)-g(r_2))|\neq0$ and $g(r_1)<g(r_2)$. Consequently, $g$ is invertible on these intervals.

Fix $\delta\in(0,1)$ and let $|u_i|_\infty=0$. Since $\id< |g|$, where $\id$ is the identity function, together with \eqref{eq:Vdot:tanh=1}, the condition
\begin{equation*}
	V_i(x_i)\geq g^{-1}\left(\dfrac{h(V_{3-i}(x_{3-i}))}{1-\delta}\right)=:\gamma_i(V_{3-i}(x_{3-i}))
\end{equation*}
implies
\begin{equation*}
	\dot{V}_i(x_i)\leq-\delta g(V_i(x_i)).
\end{equation*}
Also, for every $x_i\in(\underline{M}_k,\overline{M}_k)$, the inequality
\begin{equation*}
	\gamma_i\circ\gamma_{3-i}(V_i(x_i))<V_i(x_i)
\end{equation*}

When $n$ is infinite, the same analysis of item is employed. However, there are (countably many) intervals $(\underline{M}_k,\overline{M}_k)$. Thus, the small-gain condition holds, for the system resulting of the interconnection of \eqref{eq:subsystem:example} with $|u|_\infty=0$. Thus, Assumption~\ref{hyp:localSGC} holds.

\subsection{Proof that it satisfies the dissipation inequality in the regions where $g$ is numerically constant}

The regions where $g$ is numerically constant correspond to the closed set
\begin{equation*}
	\mathbb{R}\setminus\bigcup_{k}(\underline{M}_k,\overline{M}_k).
\end{equation*}
For every $x_i$ in this set, $\mathbin{\mathtt{round}}(1-\tanh(x_i))=0$.  Since $g$ is numerically constant, its derivative is zero. Consequently, the divergence of the vector field describing the differential equation \eqref{eq:subsystem:example} is numerically constant and equals to zero, because it is the derivative of $g$.

Consider the function
\begin{equation*}
	\begin{array}{rrcl}
		\rho:&\mathbb{R}\times\mathbb{R}&\to&\mathbb{R}_{\geq0}\\
		&(x_1,x_2)&\mapsto&e^{-(x_1+x_2)}.
	\end{array}
\end{equation*}

Since $g$ is an odd function,
\begin{align*}
	\mathbin{\mathtt{div}}(\rho f)(x):=\frac{\partial [\rho f_1]}{\partial x_1}(x)+\frac{\partial [\rho f_2]}{\partial x_2}(x)=\rho'(x)\cdot f(x)\\
	=-\sign(x_1)\rho(x_1,x_2)\bigg(-\left(25+\tfrac{|u_1|_\infty}{a+1}\right)\sign(x_1)|g(x_1)|\\
	+25h(x_2)+\left( \tfrac{|u_1|_\infty}{a+1}\right)^2\bigg)\\
	-\sign(x_2)\rho(x_1,x_2)\bigg(-\left(25+\tfrac{|u_2|_\infty}{a+1}\right)\sign(x_2)|g(x_2)|\\
	+25h(x_1)+\left( \tfrac{|u_2|_\infty}{a+1}\right)^2\bigg).
\end{align*}

Assuming $n$ finite implies that, if
\begin{enumerate}
	\item $(x_1,x_2)\in\mathbb{R}_{\leq0}^2$, then
	\begin{align*}
		\mathbin{\mathtt{div}}(\rho f)(x_1,x_2)\geq& 25\rho(x_1,x_2)\bigg(|g(x_1)|+|g(x_2)|\\
		&+h(x_1)+h(x_2)\bigg)\\
		>&0
	\end{align*}

	\item $(x_1,x_2)\in\mathbb{R}_{\geq0}^2$ and $|u_1|_\infty\cdot|u_2|_\infty\neq0$, then there exists $\varepsilon>0$ such that the inequality
	\begin{align*}
	\mathbin{\mathtt{div}}(\rho f)(x_1,x_2)
	\geq\rho(x_1,x_2)\bigg((25+\varepsilon)|g(x_1)|-25h(x_1)\\
	-\left( \tfrac{|u_1|_\infty}{a+1}\right)^2+(25+\varepsilon)|g(x_2)|-25h(x_2)-\left( \tfrac{|u_2|_\infty}{a+1}\right)^2\bigg)
\end{align*}
holds.

Since the function $\mathbb{R}\ni x_i\mapsto m_i(x_i)=(25+\varepsilon)|g(x_i)|-25h(x_i)$ is positive definite, for each index $i=1,2$, there exists a function $\gamma^{-1}\in\mathcal{K}$ such that $\gamma^{-1}(\cdot)\leq m_i(\cdot)$.

Thus, for any $\delta\in(0,1)$ the condition
\begin{equation*}
	|x_i|\geq \gamma^{-1}\left(\dfrac{1}{1-\delta}\left( \frac{|u_i|_\infty}{a+1}\right)^2\right), i=1,2
\end{equation*}
implies
$$\mathbin{\mathtt{div}}(\rho f)(x_1,x_2)>0.$$

\item $x_i<0$ and $x_{3-i}>0$, then the situation is a combination of the two previous items.
\end{enumerate}	

When $n$ is infinite or $|u_1|_\infty=|u_2|_\infty=0$, the function $\div(\rho f)$ is positive almost for almost every $x_i$ in the region where $g$ is numerically constant.

Since the above inequalities are strict and hold in a closed set, there exist a suitable open set $\mathbf{D}_k$ and a function $q_k:\mathbf{D}_k\to\mathbb{R}_{\geq0}$ satisfying Assumption \ref{hyp:a.e. dissipation inequality}.

Therefore, from Theorem \ref{thm:main result}, system resulting from the interconnection of system \eqref{eq:subsystem:example} is almost input-to-state stable.

Consider the region where $g$ is not numerically constant. In this region, there exists a subset where $\mathbin{\mathtt{div}}(\rho f)\leq0$. To see this claim, without loss of generality let $|u_1|_\infty=|u_2|_\infty=0$, and note that $\rho(0) \mathbin{\mathtt{div}}(f)(0)=-25g'(0)=-50\mathbin{\mathtt{sech}}^2(0)=-50$ and $\rho'(0)\cdot f(0)=0$. Consequently, $\mathbin{\mathtt{div}}(\rho f)(0)=-50$. From the continuity of the functions, there exists $\varepsilon>0$ such that, for every $x_1$ and $x_2\in(-\varepsilon,\varepsilon)$, the inequality $\mathbin{\mathtt{div}}(\rho f)(x)\leq0$ holds. Using the same reasoning, and due to the ``periodicity'' of the vector field $f$, the same conclusion can be obtained for every region where $g$ is not numerically constant.

\section{Conclusions and discussion}

In this paper, the interconnection of two ISS system for which a single small-gain condition does not hold everywhere on the positive real semi-axis has been considered. 

Under the assumption that there are (infinitely many) regions where the small-gain condition hold and, outside the union of these regions, the density propagation condition holds, the trajectories of solutions to the interconnected systems that do not converge to the origin have Lebesgue measure zero. An example illustrates the proposed approach.

In a future work, the authors intend to generalize the methods for the interconnection of $n>2$ ISS systems.

\printbibliography

\end{document}